\documentclass[preprint,showpacs,floats,prb]{revtex4}%
\usepackage{amsfonts}
\usepackage{float}
\usepackage{graphicx}
\usepackage{amsmath}
\usepackage{amssymb}%
\providecommand{\U}[1]{\protect\rule{.1in}{.1in}}
\begin{document}
\title{Low temperature heat capacity of Fe$_{1-x}$Ga$_{x}$ alloys with large magnetostriction}
\author{J. M. Hill}
\affiliation{Institute for Physical Research and Technology, Iowa State University, Ames, IA 50011}
\author{R. J. McQueeney}
\email{mcqueeney@ameslab.gov}
\affiliation{Department of Physics and Astronomy and Ames Laboratory, Iowa State
University, Ames, IA 50011}
\author{Ruqian Wu}
\affiliation{Department of Physics and Astronomy, University of California, Irvine, CA 92697}
\author{K. Dennis}
\author{R. W. McCallum}
\author{M. Huang}
\author{T. A. Lograsso}
\affiliation{Ames Laboratory, Ames, IA 50011}

\begin{abstract}
The low temperature heat capacity $C_{p}$ of Fe$_{1-x}$Ga$_{x}$ alloys with
large magnetostriction has been investigated. The data were analyzed in the
standard way using electron ($\gamma T$) and phonon ($\beta T^{3}$) contributions.
The Debye temperature $\Theta_{D}$ decreases approximately linearly with
increasing Ga concentration, consistent with previous resonant ultrasound
measurements and measured phonon dispersion curves. Calculations of $\Theta_{D}$ from lattice dynamical models and from measured elastic constants $C_{11}$, $C_{12}$ and $C_{44}$ are in agreement with the measured data. The linear coefficient of electronic specific heat $\gamma$ remains relatively constant as the Ga concentration increases, despite the fact that the magnetoelastic coupling increases. Band structure calculations show that this is due to the compensation of majority and minority spin states at the Fermi level.

\end{abstract}
\keywords{heat capacity, Sommerfeld constant, Debye temperature}
\pacs{75.80.+q, 65.40.Ba, 62.20.Dc, 71.15.Mb}
\received[Date: ]{August 9, 2007}

\maketitle

%corauthref\{cor\}}

\section{Introduction}

Fe$_{1-x}$Ga$_{x}$ alloys are known for their large magnetostriction. Values
of the tetragonal magnetostriction coefficient, $\frac{3}{2}\lambda_{100}$,
can reach values as high as $\sim$ 400\thinspace ppm for certain alloy
compositions and heat treatments.\cite{clark1,clark4,clark3} The sharp rise in
the magnetostriction near 19\thinspace at.\thinspace\%\thinspace Ga
composition can be explained by a simultaneously increasing magnetoelastic
coupling, $b_{1}$, and decreasing tetragonal shear modulus, $C^{\prime}%
$.\cite{clark1} The decrease in $C^{\prime}$ has been independently measured
via resonant ultrasound techniques\cite{clark1,pet,wuttig,mung1} and neutron
scattering \cite{jerel1}. However, the nature and characterization of the
large increase in magnetoelastic coupling with composition has been difficult
to determine. It has been suggested that the increase in $b_{1}$ is related to
short-range ordered clustering of the Ga atoms prior to the formation of
long-range ordered structures near\ 19\thinspace at.\thinspace\%\thinspace Ga.\cite{wu2} Below $\sim$19\%Ga, Fe$_{1-x}$Ga$_{x}$ alloys are disordered and crystallize
in a body-centered-cubic $(bcc)$ $\alpha$-Fe (A2) structure. Above this
composition, two ordered phases are possible; D0$_{3}$ and B$_{2}$. For
itinerant magnetic alloys without significant short-range ordering
(local)\ effects, $b_{1}$ depends on the spin-orbit coupling of electrons near
the Fermi level.\ In strongly ferromagnetic alloys, the compositional
dependence of the minority spin electronic density-of-states (DOS) at the Fermi
level, $n_{\downarrow}(\varepsilon_{F})$, can be related to
magnetostriction.\cite{berger77} \ Measurements of the linear coefficient of
the electronic specific heat at low temperatures, $\gamma$, (also called the
Sommerfeld constant) are directly proportional to the total electronic
DOS at the Fermi level, $n(\varepsilon_{F})$, and can be used to
characterize the origins of magnetostriction.\ This is most clearly
demonstrated in Ni-Fe alloys, where the zero magnetostriction composition
(permalloy)\ corresponds to a full majority spin band and a minimum in
$n_{\downarrow}(\varepsilon_{F})$ leading to a minimum in $\gamma$, as
predicted in the "split-band model" of Berger \textit{et al.} (Ref.\textit{
}\cite{berger77}). \ We have undertaken a study of the low tempeature heat
capacity of Fe-Ga alloys to determine if variations in $\gamma$ are present
that can be correlated with the large increase in magnetostriction.  In addition, the lattice contribution to the specific heat, characterized by the Debye temperature $\Theta_{D}$, indicates strong lattice softening with added Ga in agreement with ultrasound and neutron scattering data.

\section{Sample preparation}

Single crystal alloys of Fe$_{1-x}$Ga$_{x}$ were grown by the Bridgman
technique (see Ref.~\cite{clark4} for more details of sample preparation).
Gallium (99.999\% pure) and electrolytic iron (99.99\% pure) were cleaned and
arc melted together several times under an argon atmosphere. To prepare single
crystal samples, the as-cast ingot was placed in an alumina crucible and
heated under a vacuum to 1500 C. After reaching 1500 C, the growth chamber was
backfilled with ultra high purity argon to a pressure of 2.76 x 10$^{5}$ Pa.
Following pressurization, heating was continued until the ingot reached a
temperature of 1600 C and held for 1 hour before being withdrawn from the
furnace at a rate of 5mm/hr. Following crystal growth, the ingot was annealed
at 1000 C for 168 hours (using heating and cooling rates of 10 degrees per
minute). Small parallelepipeds (2mm $\times$1mm $\times$ 0.5mm) were cut from
the ingot by wire electrical discharge machining and cleaned by acid etching
and polished on one side. Samples were sealed in a quartz tube and annealed at
1000 C for 4 hours and furnace cooled down to room temperature. Composition
measurements were done by energy-dispersive spectrometers (EDS) in a JEOL 840A
Scanning Electron Microscope (SEM).

\section{Measurements}

\begin{figure}
[ptb]
\begin{center}
\includegraphics[
]%
{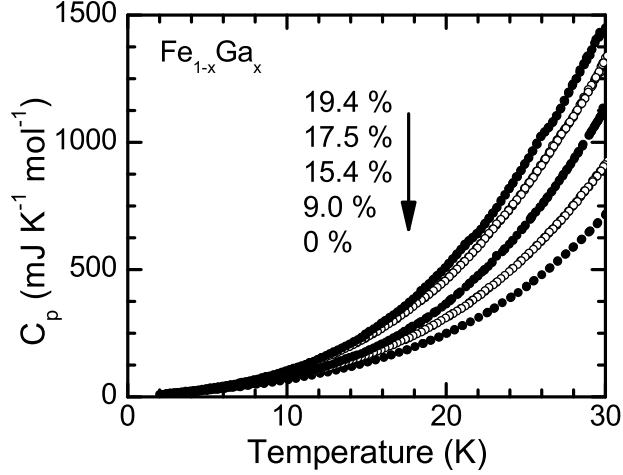}%
\caption{Heat capacity $C_{p}$ versus temperature $T$ data for slow-cooled
Fe$_{1-x}$Ga$_{x}$ alloys.}%
\label{figCp}%
\end{center}
\end{figure}

Heat capacity measurements were carried out using a Quantum Design Physical
Property Measurement System. The addenda were measured separately immediately
before the sample measurement and subsequently subtracted. \ Heat capacity
$C_{p}$ data for slow-cooled Fe$_{1-x}$Ga$_{x}$ alloys are shown in Fig.
\ref{figCp}.%
%TCIMACRO{\FRAME{ftbpFU}{4.3587in}{3.3667in}{0pt}{\Qcb{Heat capacity $C_{p}$
%versus temperature $T$ data for slow-cooled Fe$_{1-x}$Ga$_{x}$ alloys.}%
%}{\Qlb{figCp}}{fig1.eps}{\special{ language "Scientific Word";
%type "GRAPHIC";  maintain-aspect-ratio TRUE;  display "USEDEF";
%valid_file "F";  width 4.3587in;  height 3.3667in;  depth 0pt;
%original-width 4.6241in;  original-height 3.2552in;  cropleft "0";
%filename 'fig1.EPS';file-properties "XNPEU";}} }%
%BeginExpansion
%EndExpansion

\section{Data Analysis}

At low $T$, the heat capacity for a soft ferromagnet is given by the formula%
\begin{equation}
C_{p}=\gamma T+\beta T^{3}+\alpha T^{\frac{3}{2}}, \label{eqnCp}%
\end{equation}
where the terms represent the electronic, phonon, and spin-wave contributions,
respectively. The spin wave contribution has been measured at very low
temperatures and is estimated to be about a factor of fifty times smaller than
$\beta$.\cite{hather,rayne,mahesh} \ Our own analysis of the pure $\alpha$-Fe
($x=0$) data demonstrates that the heat capacity is not sensitive to the small
spin wave term proportional to $\alpha$ below 10 K.\ In this limit, the spin
wave term is ignored and the heat capacity can be written in the following
form
\begin{equation}
\frac{C_{p}}{T}=\gamma+\beta T^{2}.
\end{equation}

The plot of the data as $C_{p}/T$ vs. $T^{2}$ is shown in Fig. (\ref{figCpT2}%
). The Sommerfeld constant ($y$-intercept) and lattice specific heat
coefficient (slope) can then be obtained by a linear least-squares fit to the
plot of $C_{p}/T$ vs. $T^{2}$.\ The parameters obtained from the fits are
shown in Table~\ref{TableKey}. The Debye temperature, $\Theta_{D}$ can be
derived from $\beta$ in the procedure described below and is also shown in
Table \ref{TableKey}. The value obtained for $\gamma$ and $\Theta_{D}$ for
pure $\alpha$-Fe are consistent with literature
values.\cite{duyck,keesom,dixon,shino,landolt,aip}

\begin{center}%
%TCIMACRO{\FRAME{ftbpFU}{4.2886in}{3.6469in}{0pt}{\Qcb{Heat capacity divided by
%temperature $C_{p}/T$ versus temperature squared $T^{2}$ data for slow-cooled
%Fe$_{1-x}$Ga$_{x}$ alloys.}}{\Qlb{figCpT2}}{fig2.eps}%
%width 4.2886in;  height 3.6469in;  depth 0pt;  original-width 4.5541in;
%original-height 3.2707in;  cropleft "0";  croptop "1";  cropright "1";
%cropbottom "0";  filename 'fig2.EPS';file-properties "XNPEU";}} }%
%BeginExpansion
\begin{figure}
[ptb]
\begin{center}
\includegraphics[
]%
{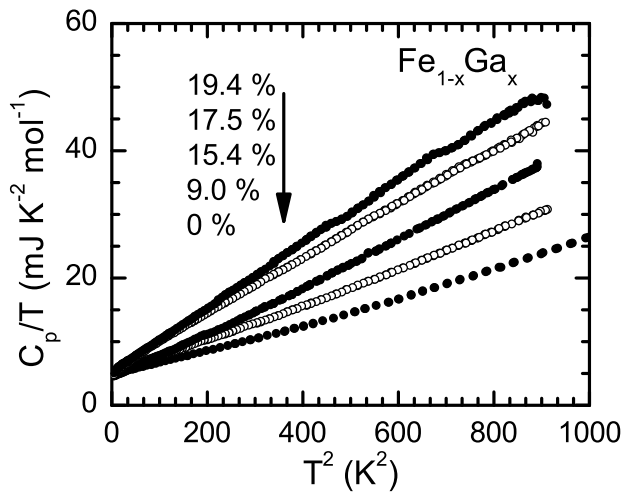}%
\caption{Heat capacity divided by temperature $C_{p}/T$ versus $T^{2}$ for slow-cooled Fe$_{1-x}$Ga$_{x}$ alloys.}%
\label{figCpT2}%
\end{center}
\end{figure}
%EndExpansion%
%TCIMACRO{\TeXButton{B}{\begin{table}[tbp] \centering}}%
%BeginExpansion
\begin{table}[tbp] \centering
%EndExpansion
\caption{Parameters obtained from fits by Eq. (2) to slow-cooled $C_{p}/T$ data.}%
\begin{tabular}
[c]{cccc}\hline\hline
at.\thinspace\% Ga & $\gamma$\thinspace($\frac{mJ}{molK^{2}}$) & $\beta
$\thinspace($\frac{mJ}{molK^{4}}$) & $\Theta_{D}$\thinspace(K)\\\hline
0.0 & 4.89(1) & 0.0187(1) & 470.2(8)\\
9.0 & 5.03(1) & 0.0276(2) & 413(3)\\
15.4 & 4.46(1) & 0.0332(1) & 388(3)\\
17.5 & 4.69(1) & 0.0582(3) & 322\\
19.4 & 5.21(1) & 0.0521(2) & 334(4)\\\hline\hline
\end{tabular}
\label{TableKey}%
%TCIMACRO{\TeXButton{E}{\end{table}}}%
%BeginExpansion
\end{table}%
%EndExpansion

\end{center}

\subsection{Electronic heat capacity}

\begin{figure}
[ptb]
\begin{center}
\includegraphics[
]%
{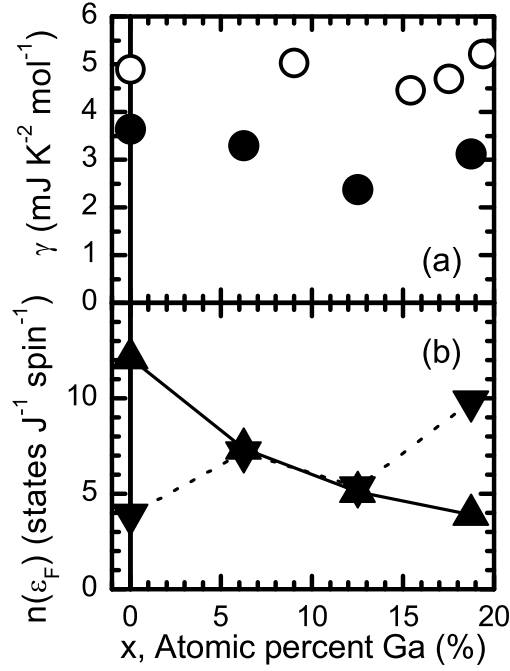}%
\caption{(a)\ The electronic coefficient of the heat capacity ($\gamma$) of
slow-cooled Fe$_{1-x}$Ga$_{x}$ alloys plotted as a function of Ga
concentration, $x$; experimental data (empty circles) and calculated (filled
circles). In experimental data, the statistical error bars are smaller than the symbol size. (b) The spin-projected electronic density of states of majority (up triangles, solid line) and minority (down triangles, dotted line)\ at the Fermi level $n(\varepsilon_{F})$. }%
\label{figgamma}%
\end{center}
\end{figure}

Figure \ref{figgamma} shows that the electronic coefficient $\gamma$ remains
essentially constant as the Ga concentration increases. \ At low temperatures,
the electronic heat capacity is proportional to the DOS at the
Fermi level $n(\varepsilon_{F})$, according to the formula $\gamma=\frac
{\pi^{2}}{3}R^{2}n(\varepsilon_{F})$, where $R$ is the universal gas constant.%

%TCIMACRO{\FRAME{ftbpFU}{4.369in}{5.3056in}{0pt}{\Qcb{(a)\ The electronic
%coefficient of the heat capacity ($\gamma$) of slow-cooled Fe$_{1-x}$Ga$_{x}$
%alloys plotted as a function of Ga concentration, $x$; experimental data
%(empty circles) and calculated (filled circles). The statistical error bars
%are smaller than the symbol size. (b) The spin-projected density of states of
%majority (up triangles) and minority (down triangles)\ at the Fermi level
%$n(\varepsilon_{F})$. }}{\Qlb{figgamma}}{fig3.eps}%
%width 4.369in;  height 5.3056in;  depth 0pt;  original-width 9.3881in;
%original-height 7.9277in;  cropleft "0";  croptop "1";  cropright "1";
%cropbottom "0";  filename 'fig3.eps';file-properties "XNPEU";}} }%
%BeginExpansion
%EndExpansion
%

%TCIMACRO{\FRAME{ftbpFU}{4.6492in}{5.4993in}{0pt}{\Qcb{The calculated
%density-of-states for Fe$_{1-x}$Ga$_{x}$ alloys with black solid lines for
%$x=0$ (i.e., pure bcc Fe), green dashed-dotted lines for $x=0.0625$, red
%dashed lines for $x=0.125,$ and blue dotted line for $x=0.1875$. The positive
%side is for the majority spin channel while the negative side is for the
%minority spin channel. Zero energy is for the position of the Fermi level.}%
%}{\Qlb{figdos}}{fig4.eps}{\special{ language "Scientific Word";
%type "GRAPHIC";  maintain-aspect-ratio TRUE;  display "USEDEF";
%valid_file "F";  width 4.6492in;  height 5.4993in;  depth 0pt;
%original-width 6.0554in;  original-height 5.4449in;  cropleft "0";
%filename 'fig4.eps';file-properties "XNPEU";}} }%
%BeginExpansion
\begin{figure}
[ptb]
\begin{center}
\includegraphics[
]%
{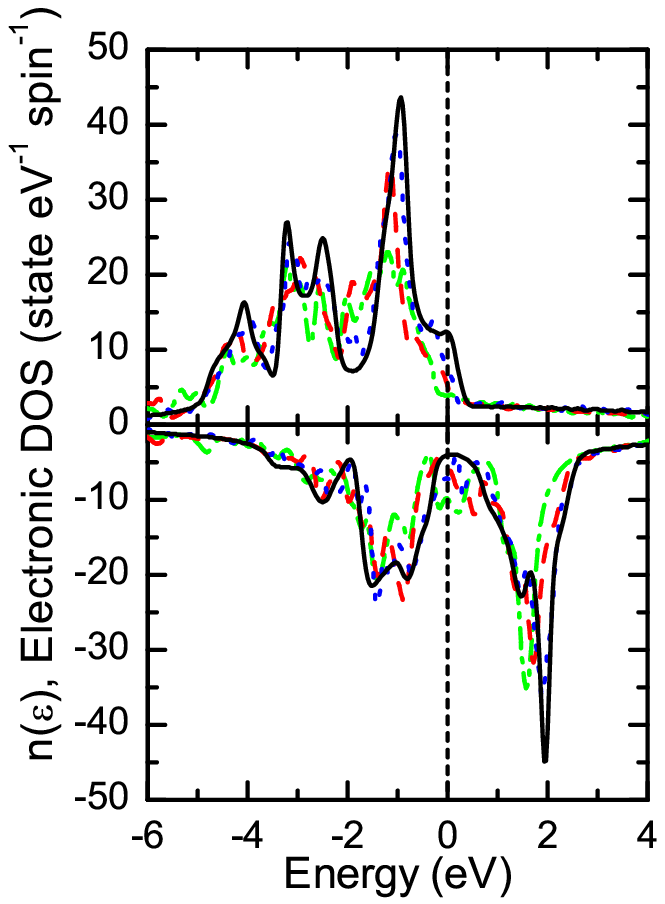}%
\caption{The calculated electronic density-of-states for Fe$_{1-x}$Ga$_{x}$ alloys with
black solid lines for $x=0$ (i.e., pure bcc Fe), green dashed-dotted lines for
$x=0.0625$, red dashed lines for $x=0.125,$ and blue dotted line for
$x=0.1875$. The positive side is for the majority spin channel while the
negative side is for the minority spin channel. Zero energy is for the
position of the Fermi level.}%
\label{figdos}%
\end{center}
\end{figure}
%EndExpansion

To better understand our experimental results, we also performed density
functional calculations for Fe$_{1-x}$Ga$_{x}$ alloys, using the highly
precise full potential linearized augment plane wave (FLAPW)
method.\cite{wimmer81} No shape approximation is assumed for the charge,
potential, and wave function expansions in the entire space. We used the
generalized gradient approximation \cite{perdew96} for the description of the
exchange correlation interaction. The convergence against parameters such as
the number of \textbf{k}-points and energy cutoff was carefully monitored. We
used a ($2\times2\times2$) supercell that comprises 16 atoms throughout the
calculations and we studied cases with $x=$ 0.0 (pure bcc $\alpha-$Fe), $x=$
0.0625 (Fe$_{15}$Ga$_{1}$), $x=$ 0.125 (Fe$_{14}$Ga$_{2}$) and $x=$ 0.1875
(Fe$_{13}$Ga$_{3}$). For the Fe$_{14}$Ga$_{2}$ and Fe$_{13}$Ga$_{3}$ cells,
there are several different ways to arrange Ga atoms on the bcc lattice sites
are related to different short-range ordered structures. Results reported
below correspond to their minimum energy configurations.  Fig. \ref{figdos}
shows the calculated electronic DOS for Fe$_{1-x}$Ga$_{x}$ alloys with $x=$ 0.0, 0.0625, 0.125 and 0.1875, respectively. As $x$ increases,
$n\left(  \varepsilon_{F}\right)  $ changes oppositely in the two spin
channels. In the majority spin part, the Fe $d$-holes are gradually purged and
$n_{\uparrow}\left(  \varepsilon_{F}\right)  $ falls monotonically. For
$x=$0.15, the Fe-$d$ band in the majority spin channel is completely
filled.\ Meanwhile, the number of non-bonding states around the Fermi level
grows steadily in the minority spin channel. The trend of the $n\left(
\varepsilon_{F}\right)  $ versus $x$ curve reasonably matches with the
experimental data of $\gamma$, as shown in Fig.\ref{figgamma}. In the
calculation, the small dip around $x=$ 0.15 mainly stems from the elimination
of the Fe majority spin $d$-holes due to the presence of Ga atoms. This dip
feature is seen also in the experimental data, however it is a small effect
that is on the limit of the sensitivity of the technique.

\subsection{Lattice heat capacity}

\begin{figure}
[ptb]
\begin{center}
\includegraphics[
]%
{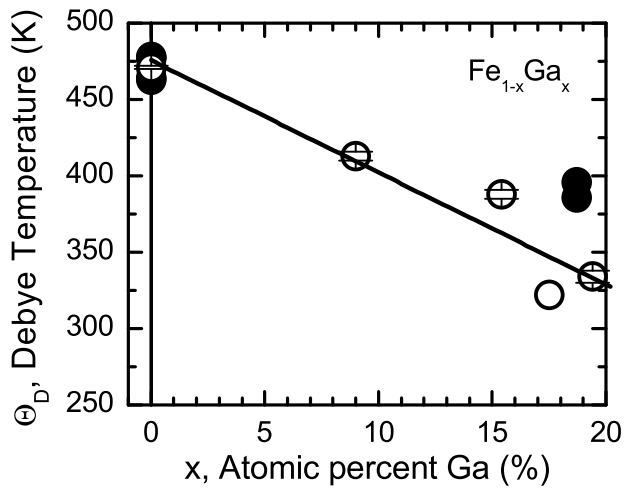}%
\caption{The Debye temperature $\Theta_{D}$ plotted as a function of Ga
concentration $x$ as determined from the data are shown as solid points
($\bullet$). The solid line is a linear fit to the slow-cooled data. Various
calculated $\Theta_{\mathrm{D}}$ values are indicated by the open circles
($\circ$), as referenced in the text.}%
\label{figdebye}%
\end{center}
\end{figure}

For a single, isotropic phonon mode with sound velocity $c$, the contribution
to the low-temperature molar heat capacity in the Debye model is
\begin{equation}
C_{V}=\frac{2\pi^{2}R}{5}\frac{k_{B}^{3}V}{\hbar^{3}c^{3}}T^{3},\label{Cv}%
\end{equation}
where $V$ is the volume of the primitive cell, $R$ is the gas constant, and
$\hbar$ is Planck's constant divided by 2$\pi$. For a general cubic crystal
with elastic anisotropy and three phonon polarizations, the form of the heat
capacity is the same as Eq.~(\ref{Cv}), with $c$ replaced by the effective
sound velocity $\bar{c}$, which is obtained by averaging the inverse-cubed
sound velocities over all possible propagation directions and modes
\begin{equation}
\frac{1}{\bar{c}\,^{3}}=\sum_{i}\int\frac{1}{\nu_{i}^{3}(\theta,\phi)}%
d\Omega.\label{ave-sound}%
\end{equation}
where $\nu_{i}(\theta,\varphi)$ is the sound velocity in a crystalline
direction given by $\theta,\varphi$. The Debye temperature is defined as
\begin{equation}
\Theta_{D}=\frac{\hbar\bar{c}}{k_{B}}\left(  \frac{6\pi^{2}}{V}\right)
^{\frac{1}{3}},\label{debyeeq}%
\end{equation}
such that the low-temperature heat capacity can be written
\begin{equation}
C_{V}=\frac{12\pi^{4}R}{5}\frac{T^{3}}{\Theta_{D}^{3}}=\beta T^{3}%
\label{phonon-heat}%
\end{equation}
and the slope of $C_{V}/T$ versus $T^{2}$ equals $\beta=$ (1943.9\thinspace J
mol$^{-1}$K$^{-1})\Theta_{D}^{-3}$. Experimentally determined values of
$\Theta_{D}$ are given in Table \ref{TableKey} and are also shown in Fig.
\ref{figdebye}. $\Theta_{D}$ decreases approximately linearly with increasing
Ga concentration, consistent with previous measurements of the tetragonal
shear modulus $C^{\prime}$ via resonant ultrasound\cite{wuttig} and neutron
scattering\cite{jerel1}.%
%TCIMACRO{\FRAME{ftbpFU}{4.3863in}{3.6167in}{0pt}{\Qcb{The Debye temperature
%$\Theta_{D}$ plotted as a function of Ga concentration $x$ as determined from
%the data are shown as solid points ($\bullet$). The solid line is a linear fit
%to the slow-cooled data. Various calculated $\Theta_{\QTR{rm}{D}}$ values are
%indicated by the open circles ($\circ$), as referenced in the text.}}%
%{\Qlb{figdebye}}{fig5.eps}{\special{ language "Scientific Word";
%type "GRAPHIC";  maintain-aspect-ratio TRUE;  display "USEDEF";
%valid_file "F";  width 4.3863in;  height 3.6167in;  depth 0pt;
%original-width 9.066in;  original-height 7.0452in;  cropleft "0";
%filename 'fig5.EPS';file-properties "XNPEU";}} }%
%BeginExpansion
%EndExpansion

In an effort to affirm the validity of the measured Debye temperatures, we
numerically calculated $\Theta_{D}$ via three methods outlined below. \ These
various estimates of $\Theta_{D}$ are also shown in fig. \ref{figdebye}.

%TCIMACRO{\FRAME{ftbpFU}{4.4598in}{7.4348in}{0pt}{\Qcb{(a) Phonon
%density-of-states $g(E)$ versus energy $E$ for $\alpha$-Fe as determined from
%a force constant model. (b) $g$($E$)/$E^{2}$ versus $E$ showing the limiting
%value (dashed line) as $E$ approaches 0\thinspace meV. (c) \ $C_{V}/T$ versus
%$T$ for the data (empty circles), as calculated from Eqn. (\ref{DOS-heat}).
%\ The best linear fit to the calculation (dashed line) gives a Debye
%temperature of 464 K.}}{\Qlb{Fe-DOS}}{fe-calc_a.eps}%
%width 4.4598in;  height 7.4348in;  depth 0pt;  original-width 5.5234in;
%original-height 9.2918in;  cropleft "0";  croptop "1";  cropright "1";
%cropbottom "0";  filename 'fig6.eps';file-properties "XNPEU";}} }%
%BeginExpansion
\begin{figure}
[ptb]
\begin{center}
\includegraphics[
]%
{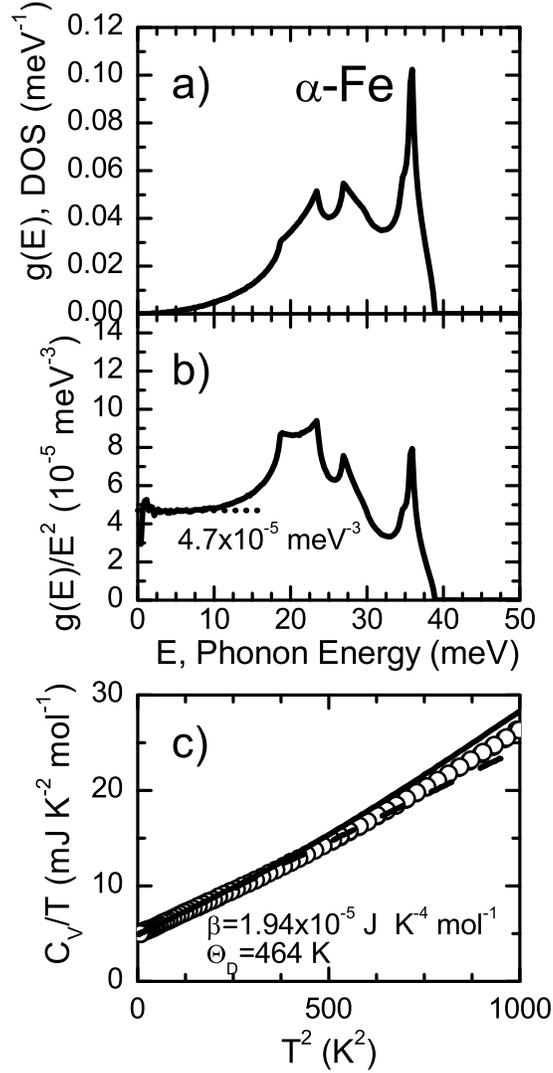}%
\caption{(a) Phonon density-of-states $g(E)$ versus energy $E$ for $\alpha$-Fe
as determined from a force constant model. (b) $g$($E$)/$E^{2}$ versus $E$
showing the limiting value (dashed line) as $E$ approaches 0\thinspace meV.
(c) \ $C_{V}/T$ versus $T$ for the data (empty circles), and as calculated from
Eqn. (\ref{DOS-heat}) (solid line). \ The best linear fit to the calculation (dashed line) gives a Debye temperature of 464 K.}%
\label{Fe-DOS}%
\end{center}
\end{figure}
%EndExpansion
\textit{(1). Integral:} \ The simplest way to calculate the Debye temperature
from a lattice dynamical model is to calculate the heat capacity by
integration. Given the model lattice dynamical parameters (obtained from fits
to inelastic neutron scattering data, for example), the phonon
density-of-states, $g(E)$, can be calculated. The phonon DOS can then be used
to calculate the heat capacity according to
\begin{equation}
C_{V}=3R\int_{0}^{\infty}g(E)\left(  \frac{E}{k_{B}T}\right)  ^{2}%
\frac{\mathrm{e}^{\frac{E}{k_{B}T}}}{\left[  \mathrm{e}^{\frac{E}{k_{B}T}%
}-1\right]  ^{2}}dE. \label{DOS-heat}%
\end{equation}
At low temperatures, the result of this calculation can be fit to the limiting
form in Eq.~(\ref{phonon-heat}) to obtain $\Theta_{D}$ in the same way as it
was obtained from the measured data above. \ Fig. \ref{Fe-DOS}(a) shows the
calculated DOS for $\alpha$-Fe where the force constants are obtained by
fitting the room temperature phonon dispersion curves measured by inelastic
neutron scattering.\cite{mink} Using this DOS, we can calculate the heat
capacity at low temperatures via Eq.~(\ref{DOS-heat}) and, subsequently, fit
the curve up to 10\thinspace K as shown in Fig.~\ref{Fe-DOS}(c). The fitted
slope, 1.94$\times10^{-5}$\thinspace Jmol$^{-1}$K$^{-4}$, can then be used to
obtain $\Theta_{D}=$ 464\thinspace K via Eq.~(\ref{phonon-heat}).

\textit{(2). Extrapolation:} \ One can obtain $\Theta_{D}$ directly from the
DOS without calculating the heat capacity. The $T^{3}$ dependence arises from
the low energy quadratic energy dependence of the DOS (resulting from a linear
dispersion relation). Plotting $\lim_{E\rightarrow0}g(E)/E^{2}$ gives the
coefficient of the low energy quadratic form of the DOS, which leads directly
to the average $\bar{c}$:
\begin{equation}
\bar{c}=\left[  \frac{V}{2\pi^{2}}\left(  \lim_{E\rightarrow0}\frac
{g(E)}{E^{2}}\right)  ^{-1}\right]  ^{\frac{1}{3}}. \label{sound}%
\end{equation}
Then, Eq.~(\ref{sound}) can be used to calculate $\Theta_{D}$.
Fig.~\ref{Fe-DOS}(b) shows $g(E)/E^{2}$ as calculated for $\alpha$-Fe. The
(average) zero frequency limit of this function has a value of approximately
4.7$\times10^{-5}$\thinspace meV$^{-3}$. Using Eqs.~(\ref{debyeeq}) and
(\ref{sound}), $\Theta_{D}$ is calculated to be 463\thinspace K.

\textit{(3) Elastic constants:} \ Additionally, the Debye temperature can be
calculated directly from the elastic constants. \ Any cubic system can be
described by three independent elastic constants; $C_{11}$, $C_{12}$ and
$C_{44}$. Generally these are experimentally measured as two transverse modes,
$C_{44}$ (rhombohedral shear) and $C^{\prime}=1/2(C_{11}-C_{12})$
(tetragonal shear), and one longitudinal mode, $K=1/3(C_{11}+2C_{12})$ (bulk
modulus). Given these elastic constants, the sound velocities $\nu_{i}(\theta
$, $\phi)$ can be calculated for any propagation direction $(\theta,\phi)$
using the Green-Cristoffel equations:%

\begin{equation}
\left\vert
\begin{array}
[c]{ccc}%
(C_{11}-C_{44})q_{x}^{2}+C{_{44}q^{2}-\rho\omega^{2}} & (C_{12}+C_{44}%
)q_{x}q_{y} & {(C_{12}+C_{44})q_{x}q_{z}}\\
(C{_{12}+C_{44})q_{y}q_{x}} & (C{_{11}-C_{44})q_{y}^{2}+C_{44}q^{2}-\rho
\omega^{2}} & (C{_{12}+C_{44})q_{y}q_{z}}\\
(C{_{12}+C_{44})q_{z}q_{x}} & (C{_{12}+C_{44})q_{z}q_{y}} & (C{_{11}%
-C_{44})q_{z}^{2}+C_{44}q^{2}-\rho\omega^{2}}%
\end{array}
\right\vert =0. \label{GC}%
\end{equation}

Solutions of Eqs.~(\ref{GC}) yield the sound velocities $\nu_{j}%
(\mathbf{q})=\omega_{j}(\mathbf{q})/q$ for each phonon branch along the
crystal direction given by wave vector $\mathbf{q}$. The average sound
velocity can then be evaluated for an anisotropic cubic crystal by numerical
averaging as shown in Eq.~(\ref{ave-sound}). Subsequently, Eq.~(\ref{debyeeq})
can be used to calculate $\Theta_{D}$.

For $\alpha$-Fe, the zero-Kelvin elastic constants are extrapolated to be
$C_{11}=$ 243.1\thinspace GPa, $C_{12}=$ 138.1\thinspace GPa and $C_{44}=$
121.9\thinspace GPa.\cite{rayne} Calculation of the Debye temperature based on
the above method gives $\Theta_{D}=$ 478\thinspace K which compares favorably
to the value determined from fits to the low temperature calorimetry data, as
shown in Fig.~\ref{figdebye}. This can also be compared to the result obtained
via the de~Launay formula; a general semi-analytic function to determine
$\Theta_{D}(T)$ for cubic metals from the elastic
constants.\cite{launay1,launay2,launay3} Given the elastic constants listed
above, this formula yields $\Theta_{D}$($T=$ 0\thinspace K) $=$ 477\thinspace
K.\cite{rayne}

The elastic constants for quenched Fe$_{81.3}$Ga$_{18.7}$ were independently
measured by Clark \emph{et~al}.\ (Ref.~\cite{clark1}) as a function of
temperature. These were extrapolated to 0\thinspace K and substituted into the
de~Launay formula yielding $\Theta_{D}$($T=$ 0\thinspace K) $=$ 386\thinspace
K. They were also used via Eqs.~(\ref{GC},~\ref{ave-sound}~and~\ref{debyeeq})
to calculate $\Theta_{D}=$ 396\thinspace K. As can be seen in
Fig.~\ref{figdebye}, these values are generally consistent with trends in the data.

\section{Discussion}

We find only a weak correlation of $\gamma$ and $\lambda_{100}$ with
composition. Band structure calculations within density functional theory show
that the small change in $\gamma$ is caused by the cancellation of two
effects; a simultaneous depletion of holes in the majority spin band and the
expected increase in $n_{\downarrow}(\varepsilon_{F})$ in the minority band.
Thus, while the heat capacity is not a sensitive probe of the relevant
electronic states for magnetostriction in the case of Fe$_{1-x}$Ga$_{x}$, the
agreement with band structure calculations is a positive step. The increase in
$n\left(  \varepsilon_{F}\right)  $ after $x>$ 0.15 manifests the presence of
non-bonding states around the Fermi level in the minority spin channel. As
revealed in our previous studies,\cite{wu2} these states play a key role for
the increase of $\lambda_{001}$ and the role of the effect of short-range
ordering is an open question. Of course, more analysis on the details of wave
functions, such as their magnetic quantum numbers, are needed for correct
prediction. Results concerning the theoretical prediction of $\lambda_{001}$
will be published elsewhere.

In addition to the electronic behavior, the low temperature heat capacity can
independently probe the lattice softening by determination of the Debye
temperature. \ We find that the decrease of the Debye temperature agrees with
previous estimates of lattice softening.

\section{Summary}

We have measured the heat capacity as a function of temperature for Fe$_{1-x}%
$Ga$_{x}$, 0.0 $<x<$ 0.194 solid solutions crystallizing in the $bcc$
structure. The Debye temperatures follow a linearly decreasing trend with
increasing Ga concentration consistent with known lattice softening. The
electronic coefficient of the specific heat remains relatively constant, in
agreement with band structure calculations.

\section{Acknowledgments}

We thank S. Bud'ko and R. Rink for performing additional heat capacity
measurements. Ames Laboratory is supported by the U.S. Department of Energy
under Contract No. DE-AC02-07CH11358. This project is partially supported by
the Office of Naval Research under ONR grant No. N000140610530.

\bibliographystyle{apsrev}
\bibliography{FeGa}

\end{document}